\documentclass[aps,prd,superscriptaddress,floatfix,nofootinbib,notitlepage]{revtex4-2}

\usepackage{aas_macros}
\usepackage{graphicx}
\usepackage{multirow}
\usepackage{amsmath,amssymb,amsfonts}
\usepackage{verbatim}
\usepackage{bm}
\usepackage{color}
\usepackage{ulem}
\usepackage{mathtools}
\usepackage{colortbl}
\usepackage[dvipsnames]{xcolor}

\usepackage[colorlinks]{hyperref}
\hypersetup{
    colorlinks=true,
    linkcolor=blue,
    filecolor=blue,      
    urlcolor=blue,
    citecolor=magenta,
    pdfpagemode=FullScreen
} 

\newcommand{\orcid}[1]{\hspace{1mm}\href{https://orcid.org/#1}{\includegraphics[height=0.3cm,keepaspectratio]{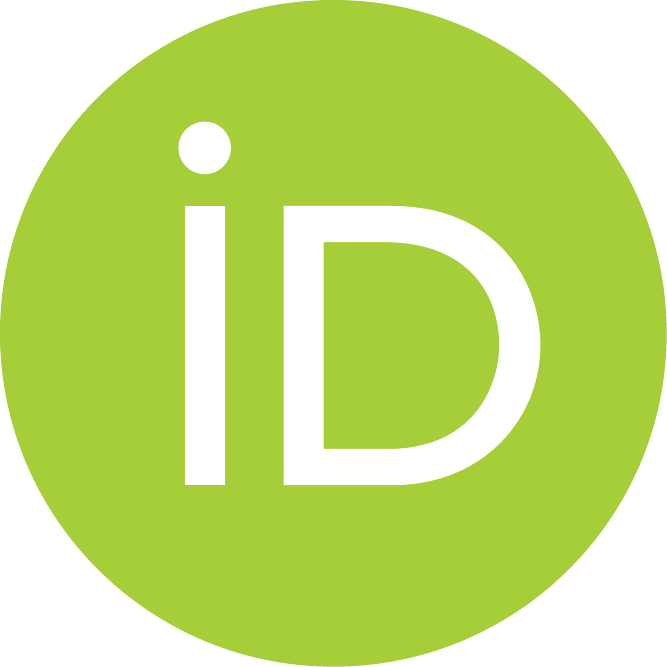}}}

\newcommand{\pprime}{{\prime\prime}}

\begin{document}

\title{TeV halos and the role of pulsar wind nebulae as sources of cosmic ray positrons}

\author{Benedikt Schroer\orcid{0000-0002-4273-9896}}
\email{bschroer@uchicago.edu}
\affiliation{Gran Sasso Science Institute (GSSI), Viale Francesco Crispi 7, 67100 L'Aquila, Italy}
\affiliation{INFN-Laboratori Nazionali del Gran Sasso (LNGS),  via G. Acitelli 22, 67100 Assergi (AQ), Italy}
\affiliation{Department of Astronomy and Astrophysics, University of Chicago, 5640 S Ellis Ave, Chicago, IL 60637, USA}

\author{Carmelo Evoli\orcid{0000-0002-6023-5253}} 
\email{carmelo.evoli@gssi.it}
\affiliation{Gran Sasso Science Institute (GSSI), Viale Francesco Crispi 7, 67100 L'Aquila, Italy}
\affiliation{INFN-Laboratori Nazionali del Gran Sasso (LNGS),  via G. Acitelli 22, 67100 Assergi (AQ), Italy}

\author{Pasquale Blasi\orcid{0000-0003-2480-599X}}
\email{pasquale.blasi@gssi.it}
\affiliation{Gran Sasso Science Institute (GSSI), Viale Francesco Crispi 7, 67100 L'Aquila, Italy}
\affiliation{INFN-Laboratori Nazionali del Gran Sasso (LNGS),  via G. Acitelli 22, 67100 Assergi (AQ), Italy}

\date{\today}

\begin{abstract}
The recent detection of the Geminga PWN by HAWC in the multi-TeV band allows us to infer precious information about the transport of pairs in the immediate surroundings of the pulsar and on the spectrum of pairs contributed by a Geminga-like pulsar to the spectrum of pairs in the cosmic radiation. Moreover, this detection allows us to address the issue of how typical are the so-called TeV halos associated to PWNe. Our calculations confirm the need to have suppressed diffusion in a region of at least $20-50\,$pc around the pulsar, and are used here to infer precious constraints on the spectrum of pairs accelerated at the termination shock: more specifically, we discuss the conditions under which such a spectrum is consistent with that typically expected in a PWN and how it gets modified once it escapes the halo. Finally, we discuss the implications of the existence of a TeV halo around Geminga in terms of acceleration of protons in the pulsar environment, a topic of profound relevance for the whole field of particle acceleration and physics of pulsars.  
\end{abstract}

\maketitle

\section{Introduction}

The detection of an extended TeV $\gamma$-ray emission around the Geminga and Monogem pulsar wind nebulae (PWNe) by HAWC \cite{geminga_hawc} and Milagro \cite{geminga_milagro} has opened many new questions with potentially crucial impact on our understanding of cosmic ray (CR) transport and the origin of~CR positrons~\cite{2022NatAs...6..199L}.  

In particular, HAWC confirmed the detection of a region of $\sim2^\circ$ around Geminga (corresponding to $\sim10$s of parsecs) in the 5-40 TeV $\gamma$-ray energy range. This emission is believed to be the result of inverse Compton scattering (ICS) of $e^\pm$ pairs accelerated at the PWN termination shock and released into the interstellar medium (ISM). Surprisingly, this interpretation requires that the diffusive transport of the pairs in the region of $20-100$ pc around the PWN occurs with a diffusion coefficient about two to three orders of magnitude lower than the average Galactic one, as inferred from measurements of cosmic ray nuclei~\cite{BC_diff,2020A&A...639A.131W}.
Alternative explanations, attempting to avoid small diffusion coefficients, have also been proposed: In~\cite{2021PhRvD.104l3017R}, a combination of ballistic and diffusive propagation was adopted to account for the observed size and TeV emission features. This explanation was questioned in~\cite{2022ApJ...936..183B} on energetic grounds as it would require a conversion efficiency from the pulsar spin-down energy to high energy leptons exceeding 100\%. In ref.~\cite{PhysRevLett.123.221103}, it was proposed that the intensity profile of Geminga's halo may reflect the smallness of the diffusion coefficient perpendicular to magnetic field lines, if the local field around Geminga happens to point toward the observer. On the other hand, the unlikely occurrence of such a situation and the fact that these models cannot simultaneously account for the size and radial symmetry of the TeV halo population have led some authors to question this approach~\cite{2022PhRvD.106l3033D}. It follows that at this time the interpretation of the TeV halos based on suppressed diffusion around PWNe remains the most plausible.

After the initial discovery, several new objects with similar properties have been found by HAWC and LHAASO \cite{LHAASO_halo}. Recently the observation of a TeV halo around a radio-quiet pulsar was claimed by HAWC \cite{radioquiet}. In ref.~\cite{halos_DM}, the authors pointed out that additional candidate TeV halo sources can be found in existing source catalogues. 
Moreover, recent observations of Geminga by H.E.S.S. \cite{geminga_hess} show the importance of having multiple experiments joining in to clarify the origin of TeV halos.

The discovery of TeV halos around PWNe raised several questions of the greatest interest: the first being the physical processes responsible for the reduced diffusivity around PWNe. It is likely that these regions are due to phenomena associated with either the PWN itself or the parent supernova in which the pulsar originated. 
Is it possible that either the escaping pairs or particles accelerated at the forward shock of the parent supernova play a role in creating turbulence that in turn confines particles for longer times? 

Second, if these regions of reduced diffusivity are confirmed and are common around PWNe, what are the implications in terms of transport of CRs in the Galaxy and more specifically about the production of CR positrons in PWNe? 
At present, there are indications that these regions of extended gamma-ray emission around galactic pulsars might be rather uncommon, with a rate of occurrence corresponding to $\sim 5\div 10$\% of the whole pulsar population, making the origin of this phenomenon even more puzzling~\cite{2020A&A...636A.113G,2022A&A...665A.132M}.


In the literature appeared so far two different hypothesis about the origin of the suppression: confinement due to outer turbulence, e.g., by the parent supernova remnant (SNR) or self-confinement by the pairs. 

The self-confinement of CRs around supernova remnants (SNRs) due to the excitation of both resonant~\cite{Malkov2013,Marta2016,Nava2016,Nava2019,2022A&A...660A..57R} and non-resonant~\cite{Schroer2021,Schroer2022} streaming instabilities has been widely investigated in the past few years and is a rather expected phenomenon. 
Resonant streaming should, however, be responsible for CR confinement only for energies $\lesssim 1$ TeV and is unlikely to be important at the energies where TeV halos are measured. 
The non-resonant streaming instability may on the other hand work for higher energy CRs, provided the CR current is strong enough. 

The excitation of a resonant instability due to pairs generated by PWNe was first discussed in~\cite{evoli_linden}.
However, in order to achieve sufficient self-confinement with the resonant streaming instability, the current needs to be spatially confined in a narrow beam  \cite{evoli_linden,linden_tevhalo}, unlikely to resemble the quasi-spherical shape observed in the case of Geminga. 
%
%
A possibility discussed in~\cite{linden_tevhalo} is that the pairs propagate in a medium that is not pristine but rather affected by the turbulence injected through the excitation of instabilities by the more abundant CR hadrons in the parent SNR. 
This scenario is hardly distinguishable from the one in which the turbulence is not CR related but rather associated with the downstream region of the parent SNR shock. Such a situation might be mimicked by adopting by hand a small correlation length $l_c$ of the assigned turbulent field~\cite{giacinti_Lc}, with $l_c\sim1\,pc$. The transport in the downstream turbulence was investigated in \cite{SNR_turbulence}, where the authors found that a Geminga-like halo could in principle arise, although no turbulence damping was introduced in the calculations.

Notice that in the case of PWNe the non-resonant instability should not be excited as, at first glance, the beam of pairs produced by the pulsar is expected to be quasi-neutral, so that the net current is vanishingly small. It has been speculated \cite{Olmi2019} that the highest energy pairs escape the PWN in a charge separated way, but it is not clear yet whether the associated electric current may be sufficient to excite the non-resonant instability to the necessary level to reproduce the suppression in diffusivity and the gamma-ray morphology observed in Geminga. 


As mentioned above, the implications of the existence of regions of reduced diffusivity around sources, and more specifically around PWNe, have profound implications for the origin of CR positrons. The positron excess, discovered by Pamela~\cite{pamela_positrons} and confirmed by FERMI~\cite{2012PhRvL.108a1103A}, and AMS-02~\cite{Ams_positrons}, clearly indicated that sources of primary CR positrons must exist in the Galaxy. Candidate sources of such positrons abound, from pulsars \citep{pulsar_primary_hooper,linden_pulsar_primary,pulsar_primary_and_DM,pulsar_primary_DiM} to mature SNRs \citep{SNR_e+_PB,SNR_e+_Ahlers,SNR_e+_Tomassetti,SNR_e+_cholis,SNR_e+_PM} and dark matter \citep{DM_decay_e+,pulsar_primary_and_DM}, but pulsars are considered to be the most reliable possibility, because there is independent evidence of the production of pairs in these sources, based on their multi-frequency appearance~(e.g.,\cite{2020A&A...640A..76P}). 
The discovery of TeV halos around PWNe is the most striking evidence that these sources liberate high-energy pairs into the ISM. The question is then whether the spectrum and fluxes of these pairs are appropriate to describe the positron flux observed at the Earth~\cite{2022A&A...665A.132M}.  


In the discovery paper by the HAWC collaboration \cite{geminga_hawc}, it was argued that if the diffusion coefficient inferred around Geminga were to be assumed to be representative of the diffusion coefficient throughout the Galaxy, then nearby PWNe such as Geminga could not contribute to the positron flux at the Earth, because of the severe energy losses and long propagation times. However, the assumption of a small diffusion coefficient in the Galaxy would contradict observations of secondary nuclei, while a more reasonable starting point is the existence of regions of reduced diffusivity, as discussed above. In this case, the contribution of nearby PWN may be appreciable \cite{tevhalo_convection,e+fraction_geminga_fang,tang_piran,fermi_halo}. 


In most previous studies of the role of PWNe to the CR positron spectrum at the Earth, the pair spectrum was assumed to be a simple power law with an index of $1.8-2.3$. with steeper spectra leading to requiring unreasonably high efficiencies ($\gtrsim 100\%$ of the total spin down power of the pulsar) to fit the data \citep{fermi_halo}.

The spectrum of pairs in PWNe as inferred from multi-frequency observations of different nebulae is a broken power with a break at lepton energy $\sim 100-1000$ GeV. This more realistic scenario was recently employed in \cite{e_factories} to make an assessment of the role of PWNe in their bow shock phase to the CR positron flux. While the efficiency required in this investigation is around $10-20\%$, the slope of the high energy spectrum, $\sim 2.8$, is somewhat steeper than what direct observation of PWNe hint at ($\sim 2.3-2.5$); a finding that was quoted in \cite{e_factories} as a source of tension. Similar efficiencies were quoted in a recent investigation including the effect of the low diffusivity regions around PWNe \cite{halo_population_study}, described using a two-zone model with a pure power-law spectrum of pairs from PWNe, with slopes in the range $1.4-2.2$. This study showed that the results are rather sensitive to the size of the halos around PWNe, typically in the range $\sim30-60\,$pc \citep{halos_DM,tang_piran}. In order to have no significant impact on nuclear secondary to primary ratios, such as $B/C$, it was estimated that the halos' size should not exceed $\sim 50\,$pc \cite{tevhalo_convection}. 

In the present article, we discuss the issue of the lepton transport in the regions of reduced diffusivity and the positron contribution to the spectrum measured at the Earth, with some noteworthy advancements with respect to previous studies: 1) we account for spectra of the pairs at the PWN that are consistent with the observations of radiation from individual nebulae (broken power laws) and we investigate the role of losses for different sizes of the halos and different strengths of the magnetic field in the same regions. 2) We calculate the spectrum and morphology of the gamma-ray emission from the halo around Geminga and how this is affected by the size of the region of reduced diffusivity and the strength of the magnetic field. In this way, we estimate the minimum size of the halo compatible with observations. 3) For the first time, we calculated the contribution of protons extracted from the neutron star surface and propagated in the region of low diffusivity to the gamma-ray spectrum, a result that will become even more important when higher energy measurements of the flux will become available. 4) We use a corrected two-zone model compared to the one adopted in previous investigations: such solutions were erroneously written in a form that does not conserve the flux at the boundary between the halo and the ordinary ISM \cite{Osipov}. 5) We calculate the effective spectrum of pairs escaping the halo and injected into the ISM, and find that the escape spectrum is severely affected by energy losses during transport in the regions of reduced diffusivity. We use this argument to establish contact with a previous investigation we carried out in \cite{e_factories}, suggesting that such modifications are necessary. 6) Finally we used the correct two-zone model of CR positron transport to estimate the contribution of a Geminga-like nearby PWN to the CR positron spectrum at the Earth.

The paper is structured as follows: in section \ref{sec:theory}, we discuss the formalism used to describe particle transport in the region around PWNe and the spectra of pairs and protons contributed by an individual PWN. In section \ref{sec:gamma}, we summarize the main contributions to the gamma-ray emission from both leptons and hadrons. In section \ref{sec:results}, we discuss our results in terms of morphology of the gamma-ray emission, spatially integrated spectrum, role of protons generated in the PWN and, most important, modification induced by energy losses to the spectrum of pairs escaping the halo around the PWN and implications for the positron spectrum at the Earth. We summarize and conclude in section \ref{sec:conclusion}.


\section{Pair transport around PWNe}
\label{sec:theory}

The number density of electron-positron pairs, $n(E,r,t)$ with energy $E$, at distance $r$ from the PWN (assumed to be at $r=0$) and at time $t$ is described by the following transport equation:

%
\begin{equation}\label{eq:transport}
\frac{\partial n(E,r,t)}{\partial t} = \frac{1}{r^2}\partial_r [ r^2 D(E,r) \partial_r n(E,r,t)] + \partial_E \left[ b(E) n(E,r,t)\right] + Q(E,r,t),
\end{equation}
where the first term on the right-hand-side describes diffusion, the second one describes energy losses due to IC (Inverse Compton) and synchrotron losses with $b(E) = -\frac{dE}{dt}$ and $Q(E,r,t)$ denotes the time-dependent release of pairs at the location of the PWN. At the high energies of interest here, other channels of energy losses, such as ionization and adiabatic losses can safely be neglected. 
The geometry of the problem is assumed to be spherically symmetric and the injection is taken to be as a $\delta$-function in space which is justified given the small spatial extent of the nebula $\sim 0.1$~pc~\cite{2017ApJ...835...66P} with respect to the overall region that we aim at describing, $\sim 50\,$pc.

In the literature, two different scenarios are considered: a) a spatially constant diffusion coefficient with boundary at infinity, b) the so-called ``two-zone model'', where the diffusion coefficient near the source (region 1) is different and smaller than the one outside (region 2). 
Since scenario a) is in conflict with CR nuclei data, all our results are obtained by adopting scenario b).

In both cases, for the energies we are interested in, the role of energy losses is dominant and the escape of pairs from the Galaxy, usually modelled in terms of a free escape boundary condition, can be safely neglected. Within this assumption, the solution of eq.~\ref{eq:transport} can be found analytically and takes the form:
\begin{equation}\label{eq:greens}
n(E,r,t) = \frac{1}{b(E)} \int_0^{t-t_{\rm BS}} \mathrm{d}t^\prime \, b(E^\prime) Q(E^\prime,t-t^\prime) \mathcal H(r,E^\prime,t^\prime),
\end{equation}
where $E^\prime$ is obtained by inverting $t(E) = \int_E^{E^\prime}\mathrm{d}E^\pprime / b(E^\pprime)$ and corresponds to the initial energy a particle of energy $E$ had a time $t$ ago because of the energy losses.
Note that the integration is carried out over the time that has passed after the particles were injected $t^\prime$. So that $t^\prime=t-t_{\rm BS}$ corresponds to particles injected at $t_{\rm BS}$ and $0$ corresponds to particles that are injected now at time $t$. The solution is valid for $t>t_{\rm BS}$.


The radiation fields adopted for the computation of energy losses are the ones from model 2 in \cite{isrf_model_delahaye} which are parametrized as blackbody or grey body radiation fields consisting of the CMB, an IR, a stellar (optical) and three UV components.
The numerical values of the temperatures and energy densities of the different components can be found in Table 2 of \cite{isrf_model_delahaye}. 
These fields are obtained by fitting the ISRF in the solar proximity obtained from the spatial dependent ISRF model of~\cite{porter_isrf}.
The magnetic field to compute the losses due to synchrotron emission is more uncertain and we leave it as a free parameter of the model as discussed in the next Section. Here, there is a subtle point to keep in mind: the analytical solution of the transport equation used here (see below) is limited to the case in which energy losses are the same everywhere in the diffusion model (even in the two zone model of diffusion). Hence, the value of the magnetic field adopted as a free parameter, which affects synchrotron losses, must be the same everywhere. This needs to be kept in mind in the cases in which the magnetic field adopted in the near-source region is much smaller than the one typically adopted in the Galactic disc. 

In eq.~\ref{eq:greens}, we take into account that PWNe inject e$^\pm$-pairs into the ISM only after they leave the parent SNR, at a time $t_{\rm BS}$ in which they develop a bowshock structure. This time depends on several parameters such as the pulsar birth kick velocity and the explosion energy of the parent remnant.
To simplify our analysis we fix this parameter to $56\,$kyr as in~\cite{Evoli2020arxiv}. 
Since we focus on the TeV emission where the loss time is much shorter than the age of Geminga, the results are basically independent of $t_{\rm BS}$.

In the case of a spatially constant diffusion coefficient, parametrized as $D = D_0 (E/100\,{\rm TeV})^\delta$, the function $\mathcal H(r,E,t)$ simply reads:
\begin{equation}\label{eq:one_zone}
\mathcal H(r,E,t) = \frac{1}{[4\pi \lambda^2(E,E^\prime)]^{3/2}} {\rm e}^{-\frac{r^2}{4\lambda^2(E,E^\prime)}},
\end{equation}
where the loss length $\lambda$ is defined in such a way that  $\lambda^2(E,E^\prime) = \int_E^{E^\prime}\mathrm{d}E^\pprime \frac{D(E^\pprime)}{b(E^\pprime)}$. The quantity $\lambda$ is the distance covered by pairs under the action of diffusion and radiative cooling from energy $E^\prime$ to $E$.

In the two-zone model, in which the diffusion coefficient rises from a value $D_0$ in the near source region to a value $D_1 > D_0$ at $r_0$, we find the following expression for the 
$\mathcal H$:
\begin{equation}\label{eq:twozone}
\mathcal H(r,E,t) = \int_0^{\infty}\mathrm{d}\psi \frac{\xi e^{-\psi}}{\pi^2\lambda_0^2 [A^2(\psi)+B^2(\psi)]}
\begin{cases}
r^{-1} \sin\left( 2\sqrt{\psi}\frac{r}{\lambda_0} \right) & 0 < r < r_0 \\
A(\psi) r^{-1} \sin(2\sqrt{\psi}\frac{r \xi}{\lambda_0}) + B(\psi) r^{-1} \cos(2\sqrt{\psi}\frac{r \xi}{\lambda_0}) & r \geq r_0, \\
\end{cases}
\end{equation}
with
\begin{equation}
A(\psi) = \xi \cos\left( \chi \right) \cos\left( \xi \chi \right) + \sin\left( \chi \right) \sin\left( \xi \chi \right)
+ \frac{1}{\chi} \left(\frac{1-\xi^2}{\xi} \sin( \chi ) \cos( \xi \chi) \right),
\end{equation}
and 
\begin{equation}
B(\psi) = \frac{\sin\left( \chi \right)-A(\psi)\sin\left( \xi \chi \right)}{\cos\left( \xi \chi \right)}.
\end{equation}
Here we introduced the quantity $\chi = 2\sqrt{\psi}\frac{r_0}{\lambda_0}$ and $\lambda_0$ is defined as above, using the diffusion coefficient $D_0$, and $\xi=\sqrt{D_0/D_1}$. 

This solution was recently derived by ~\cite{Osipov}, where the authors stress that it profoundly differs from a solution \cite[]{tang_piran} that has been previously used in the literature (see for instance \citep{halo_population_study,fermi_halo}). 
In particular, the solution of \cite{tang_piran} does not conserve the flux at the boundary between the two zones and is therefore incorrect. 
However, it is worth emphasizing that for $r_0/\lambda_0 \rightarrow \infty$ or $\xi=1$ both simplify to the usual one zone model solution given in eq.~\ref{eq:one_zone}. 
The most prominent quantitative differences between the solution in Eq. \ref{eq:twozone} and that of \cite{tang_piran} appear when $r_0 \sim \lambda_0$, which might be a situation of interest for TeV halos.
Additional differences between the two solutions are discussed in more detail in \cite{Osipov}.

%
%

%
%

The injection term $Q(E,t)$ is modelled following~\cite{BlasiAmato2011,injectionmodel_ab} as a continuous injection in time.
The observed spectrum of particles released by PWNe is well described by a broken power law~\cite{2011MNRAS.410..381B}:
\begin{equation} \label{eq:injection}
Q(E,t) =  Q_0(t) {\rm e}^{-\frac{E}{E_c(t)}}
\begin{cases}
\left(\frac{E}{E_b}\right)^{-\gamma_L} & E < E_b \\
\left(\frac{E}{E_b}\right)^{-\gamma_H} & E \ge E_b, \\
\end{cases}
\end{equation}
where typical values for the slopes below and above the break are $\gamma_L\approx 1.5-1.8$ and $\gamma_H\approx 2.2-2.8$, respectively, and $E_b\approx 100-1000\,$GeV.
The cutoff energy $E_c(t)$ is determined by the potential drop of the given pulsar at hand. 
Its present-day value can be accurately calculated using the spin-down luminosity, as $E_c \simeq 1.7 \sqrt{L_{36}}\,$PeV where $L_{36}$ is the observed spin-down luminosity today in units of $10^{36}$erg/s.

%
%

The normalization of the spectrum $Q_0(t)$ is provided by the condition that at any given time a fixed fraction $\eta$ of the spin-down luminosity $L(t)$ is converted into pairs (or other particles, if any are produced by the pulsar):
\begin{equation}
\eta L(t) = \int_{E_ {\rm min}}^\infty \mathrm{d}E E Q(E,t) \, .
\end{equation}

The time dependence of $L(t)$ is assumed to be well described by a situation in which the pulsar spin-down energy is dissipated via magnetic dipole radiation (braking index of $n = 3$). 
The braking index has been measured only in young objects, often finding values different from $3$, while for Geminga (which is a relatively old pulsar) there are no available measurements, hence we adopt the reference value of $n=3$.
We notice that the energy loss time of electrons and positrons at the energies we are interested in is $\sim 10$~kyrs, hence the results depend only weakly on the overall time evolution of the luminosity and therefore on the braking index.

The spin-down luminosity can then be written as 
\begin{equation}
L(t)=L_0\frac{(1+t_{age}/\tau_0)^2}{(1+t/\tau_0)^2}
\end{equation}
where $L_0 = 3.26\times 10^{34}\,$erg/s is the measured value today, $t_{\rm age}=342\,$kyrs is the inferred age and $\tau_0 = 12\,$kyrs is the spin-down timescale of Geminga \cite{geminga_hawc}.

\section{$\gamma-$ray emission from pairs and protons} \label{sec:gamma}

The pairs that leave the PWN into the surrounding ISM produce gamma radiation mainly by ICS the photons of the  Interstellar Radiation Field (ISRF), while X-rays can be produced in the form of synchrotron emission.

For the ICS emission we adopt the formalism introduced in \cite{khangulyan_IC}, which accounts for both the Thomson and the Klein-Nishina regimes and the transition between the two. This is especially important in our calculations given the high energies of the electrons responsible for the gamma-ray emission. 

The rate of $\gamma-$rays per unit frequency $\nu$ of an electron up-scattering an isotropic gray-body distribution with temperature $T$ can be written as
\begin{equation}\label{eq:cross_iso}
\frac{dN_{\gamma}}{d \nu\,dt} = 
{4e^4m_{\rm e}\kappa (k_{\rm B}T)^2\over \hbar^2E^2} \times
\left[{z^2\over2(1-z)}F_3(x_0)+F_4(x_0)\right] \,,
\end{equation}
where $k_{\rm B}$ is the Boltzmann constant, $\hbar$ the Planck constant, $z=h\nu /E_e$, $x_0=zm_ec^2/(4(1-z)\gamma_e k_B T)$, $\kappa$ the grey body dilution factor, and $F_3$ and $F_4$ are known functions (see eq. 24 of ref.~\cite{khangulyan_IC}).

The emissivity from ICS emission by a given distribution of electrons $n(E,r,t)$ is obtained as:
\begin{equation}\label{eq:flux}
\phi_\gamma(E_\gamma,r) = \frac{1}{4 \pi d^2} \int_0^{E_{max}}\mathrm{d}E_e n(E_e,r,t) \frac{dN_\gamma}{hd\nu dt}(E_e)\\
\end{equation}
with the distance of the source $d$.

Although the bulk of the gamma-ray emission from the region around the PWN is expected to be produced by the copious ICS emission from pairs, at least at the highest energies, it is possible that protons extracted from the neutron star surface may provide a non-negligible contribution, through the production and decays of neutral pions. 

The acceleration of protons at the termination shock was discussed by \cite{AmatoArons}, and the possibility that protons and nuclei may be energized in the magnetosphere of young rapidly spinning neutron stars, and possibly contribute to the flux of ultra high energy cosmic rays at the Earth, was first put forward in \cite{BlasiEpsteinOlinto,Arons2003}.

It is therefore meaningful to ask ourselves if there is a possible gamma-ray signature of accelerated protons to the TeV gamma-ray emission from the regions around PWNe.



The spectrum of protons (and nuclei) from PWNe is expected to be very hard, mainly depending on the braking index $n$, so that, if any effect is to be expected, it should be at high energies.

Clearly protons do not suffer severe energy losses in the region surrounding PWNe, contrary to what happens to electron-positron pairs, one more reason for expecting their potential contribution to the highest energy gamma ray emission.


The injection term for protons at the location of the PWN is assumed in the form proposed in \cite{pulsar_protons_olinto} and \cite{pulsar_protons_kotera} although we generalized the results to an arbitrary braking index $n$, which regulates the rate of neutron star spin down $\Omega=\Omega_0 (1+\tau/\tau_0)^{-1/(n-1)}$ with time. As discussed above, the default case is that of a magnetic dipole corresponding to $n=3$.

Within this model the proton injection spectrum reads:
\begin{equation}\label{eq:proton_injection}
    Q_p(E_p,t) = \eta_p \dot{N}_{GJ}(t) \delta(E_p-E_c(t))\,,
\end{equation}
where $E_c(t)$ is the same potential drop as in the case of the pairs at present time, scaling as $\Omega^{2}$, the injection efficiency of protons is $\eta_p$ and $\dot{N}_{GJ}(t)=\sqrt{L_0c}e^{-1} (\Omega/\Omega_0)^2$ is the Golreich-Julian-density. Here $e$ is the unit charge and $L_0$ the spin-down luminosity.

The injected spectrum, i.e., the time integral of eq. \ref{eq:proton_injection}, scales as $E^{-(n-1)/2}$, recovering the $E^{-1}$ spectrum of \cite{pulsar_protons_olinto} and \cite{pulsar_protons_kotera} for $n=3$ and even harder spectra for lower $n$.

Since electromagnetic losses are negligible for protons, we solve equation \ref{eq:greens} for $b(E)=0$, using the same diffusion coefficient adopted for the transport of pairs. The corresponding solution of the transport equation for protons is again in the form of eq. \ref{eq:twozone}
\begin{equation}
    n_p(E_p,r,t) = \int^\infty_0\mathrm{d}t_0 Q_p(E_p,t_0) \mathcal H(r,E_p,t-t_0)\,,
\end{equation}
where now $\lambda_0=2\sqrt{D(E) t}$ in $\mathcal H(r,E_p,t-t_0)$. Here, we assume that all protons escape the nebula even at $t<t_{\rm BS}$, so as to maximize the flux of protons injected by the pulsar. This choice ensures that the obtained flux is the absolute maximum that one can extract. As we will discuss later, even such an extreme choice leads to a subdominant role of protons in the Geminga TeV halo.

After escaping the PWN, protons interact with the surrounding ISM gas thereby producing neutral pions, which in turn decay into gamma rays. The emissivity in the form of gamma radiation is calculated using the formalism introduced by \cite{kelneraharonian}: 
\begin{equation}
\Phi_\gamma(E_\gamma,r)
= c\,n_{H}^{}\int\limits_{E_\gamma}^\infty\! \sigma_{\rm inel}(E_p)\,
n_p(E_p,r,t)\,F_\gamma\left(\frac{E_\gamma}{E_p},\,E_p\right)\frac{dE_p}{E_p}\,
\end{equation}
where the function $F_\gamma\!\left(\frac{E_\gamma}{E_p},\,E_p\right)$ is provided in \cite{kelneraharonian}, $n_H$ is the ambient gas number density and $\sigma_{\rm inel}(E)$ is the inelastic cross section of p-p scattering.

In Sec. \ref{sec:protons}, we will discuss the calculations of the gamma-ray emission from pp scattering, using the typical gas density of the ISM $n_H=1\,$cm$^{-3}$ and an injection efficiency of protons $\eta_p=1$. Clearly, these conditions provide an upper limit to the contribution of protons to the gamma-ray emission of the TeV halo around Geminga, since the efficiency is bound to be smaller than unity and X-ray measurements in the circum-PWN region suggest that the density of gas might be $\sim 0.01\,$cm$^{-3}$ in the immediate vicinity of Geminga \cite{posselt}.

To compute the total $\gamma$-ray flux $\Phi_\gamma$ and compare it with the one measured by HAWC, we integrate over the LOS and the whole field of view used for the HAWC measurement. The field of view is approximated as a circle of radius $\rho_{max}=44\,$pc around Geminga, taken from \cite{geminga_hawc}, while the LOS are approximated as being parallel to each other resulting in
\begin{equation}\label{eq:total_gamma_flux}
    \Phi_\gamma(E) = 2\pi \int_{-d}^\infty\mathrm{d}l \int_0^{\rho_{max}}\mathrm{d}\rho \rho \phi_{\gamma}(E_\gamma,\rho,l)\,,
\end{equation}
where we changed from the spherical radius $r$ centered around the pulsar to cylindrical coordinates $\rho$ and $l$ .
The surface brightness measured by HAWC is obtained as the LOS integral of the obtained $\gamma$-ray emissivities $\phi_\gamma(E_\gamma,r)$, corresponding to the integral over $l$ in eq.~\ref{eq:total_gamma_flux}.



\section{Results}\label{sec:results}

Given the complexity of the problem and the numerous parameters that are necessary to describe it, we fix the parameters listed in table \ref{tab:default_parameters}, and we investigate the effect of changing certain parameters, in terms of spatial morphology and total flux of the gamma-ray emission. 


In table \ref{tab:default_parameters}, the source age and distance are chosen following \cite{geminga_hawc}.
The break position in the spectrum of pairs released by the PWN is fixed at the rather high energy of $1\,$TeV, motivated by the study of \cite{posselt} where a rather hard spectrum is found for particles of a few hundred GeV, suggesting that if a break is present it has to be at a rather high energy. Note however, that the exact break position only affects the efficiency needed to reproduce the HAWC measurements, with a break at higher energies requiring lower efficiencies.

In table~\ref{tab:cases}, we present the different cases that we study in the following together with their values for the high-energy injection slope, magnetic field, diffusion coefficient, halo size and injection efficiency.
We write the diffusion coefficient in the region around the PWN in multiples of the best-fit value found by HAWC in ref.~\cite{geminga_hawc} $D_{\text{HAWC}}= 3.2 \cdot 10^{27}\,$cm$^2$/s.
 This diffusion coefficient is suppressed by a factor $\sim 1300$ compared to the average Galactic diffusion coefficient obtained from fits to secondary to primary ratios in \cite{BC_diff} at $100\,$TeV energies.


\begin{table}[]
    \centering
    \begin{tabular}{c|c|c|c|c|c|c}
    $t_{age}$  & distance   &  $E_b$ & $\gamma_L$ & $\gamma_H$ & $B$ & $D_0$\\
    \hline 
    $342\,$kyr & $250\,$pc    & $1000\,$GeV & $1.5$ & $2.5$ & $3\,\mu$G & $3.2\cdot 10^{27}$cm$^2$s$^{-1}$
    \end{tabular}
    \caption{Default values for our parameters unless otherwise stated in each scenario they are fixed to these values.}
    \label{tab:default_parameters}
\end{table}

\begin{table}[]
    \centering
    \begin{tabular}{c|c|c|c|c|c|c}
    case & $\gamma_H$ & $B$ & $D_0$ & $r_0$ & $\epsilon$\\
    \hline 
    I & $2$ & $3\,\mu$G & $D_{\text{HAWC}}$ & $50\,$pc & $0.09$\\
    II & $2$ & $1\,\mu$G & $0.5\cdot D_{\text{HAWC}}$ & $50\,$pc & $0.04$\\
    III & $2$ & $6\,\mu$G & $3\cdot D_{\text{HAWC}}$ & $50\,$pc & $0.26$\\
    IV & $2$ & $3\,\mu$G & $D_{\text{HAWC}}$ & $20\,$pc & $0.14$\\
    V & $2$ & $3\,\mu$G & $10 \cdot D_{\text{HAWC}}$ & $20\,$pc & $0.83$\\
    VI & $2.3$ & $3\,\mu$G & $D_{\text{HAWC}}$ & $50\,$pc & $0.20$\\
    VII & $2.5$ & $3\,\mu$G & $D_{\text{HAWC}}$ & $50\,$pc & $0.37$\\
    VIII & $2.3$ & $1\,\mu$G & $0.5\cdot D_{\text{HAWC}}$ & $50\,$pc & $0.09$\\
    \end{tabular}
    \caption{Compilation of different parameter combinations discussed in this article.}
    \label{tab:cases}
\end{table}

\subsection{Spatial Morphology}

The main piece of observation that leads to infer a small diffusion coefficient around the Geminga PWN is the gamma-ray morphology as measured by HAWC. However, it is important to realize that the strength of the suppression in the diffusivity depends on other parameters as well, and most important on the strength of the magnetic field in the same region. 
In fact, rather than constraining directly the diffusion coefficient, the gamma-ray morphology allows one to infer the loss length $\lambda_{50TeV}\approx \sqrt{D\tau_{loss}}$, namely the distance covered by electrons of given energy in one loss time. The loss time is however determined by the magnetic field strength (because of synchrotron losses) and ICS, in a region where the Klein-Nishina correction is not negligible. Increasing the strength of the magnetic field leads to a decrease in the loss time, so that larger diffusion coefficients are allowed. The opposite happens when the magnetic field is decreased. 

\begin{figure}
    \centering
    \includegraphics[width=0.5\textwidth]{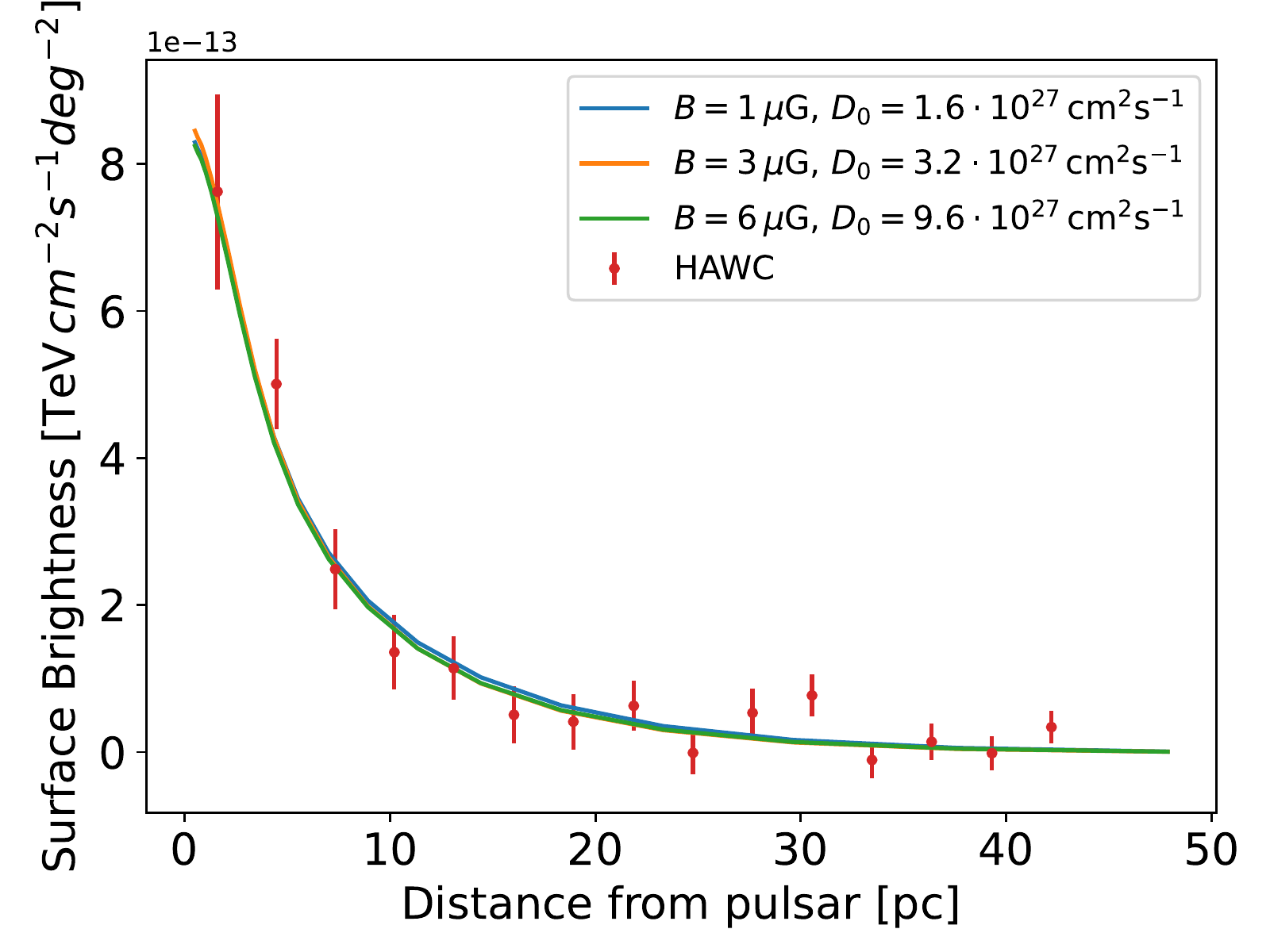}
    \caption{$\gamma$-ray spectrum produced by the released particle spectra for different combinations of $B$ and $D$ in order to keep $\lambda_{50TeV}$ constant. The used efficiencies are $0.04$, $0.09$ and $0.26$ with increasing magnetic field strength.}
    \label{fig:morph_diff_B}
\end{figure}

In order to illustrate this point better, in fig. \ref{fig:morph_diff_B}, we show the spatial morphology obtained with different choices for the magnetic field strength $1$, $3$, and $6\,\mu$G and correspondingly different diffusion coefficients, chosen in such a way as to keep $\lambda_{50TeV}\approx 11.9\,$pc, see cases I-III in table~\ref{tab:cases}.

It is evident that the three spatial profiles are practically identical and describe well the HAWC observation, although the diffusion coefficient and the injection efficiencies differ by a factor six in the different cases. For $\gamma_H=2$ the inferred efficiency values are $0.04$, $0.09$ and $0.26$ with increasing magnetic field strength respectively because larger magnetic field means larger diffusion coefficient and therefore larger required efficiency.

This degeneracy should be kept in mind when addressing the issue of the physical origin of the suppressed diffusion based on the region of gamma-ray emission, as changing the magnetic field strength can change the magnitude of the suppression by a factor of a few.


Fig. \ref{fig:morph_diff_B} also explains why the gamma-ray morphology is only weakly dependent upon the energy dependence of the diffusion coefficient (for instance Kraichnan versus Kolmogorov scaling) as long as the normalization of $D$ at $\sim50\,$TeV remains approximately the same. In the extreme case of no suppression of the diffusion coefficient, the observed spatial morphology would require a magnetic field of $\sim 100\,\mu$G, a configuration that cannot be realized since it would require an efficiency of injection of pairs exceeding unity.
\begin{figure}
    \centering
    \includegraphics[width=0.5\textwidth]{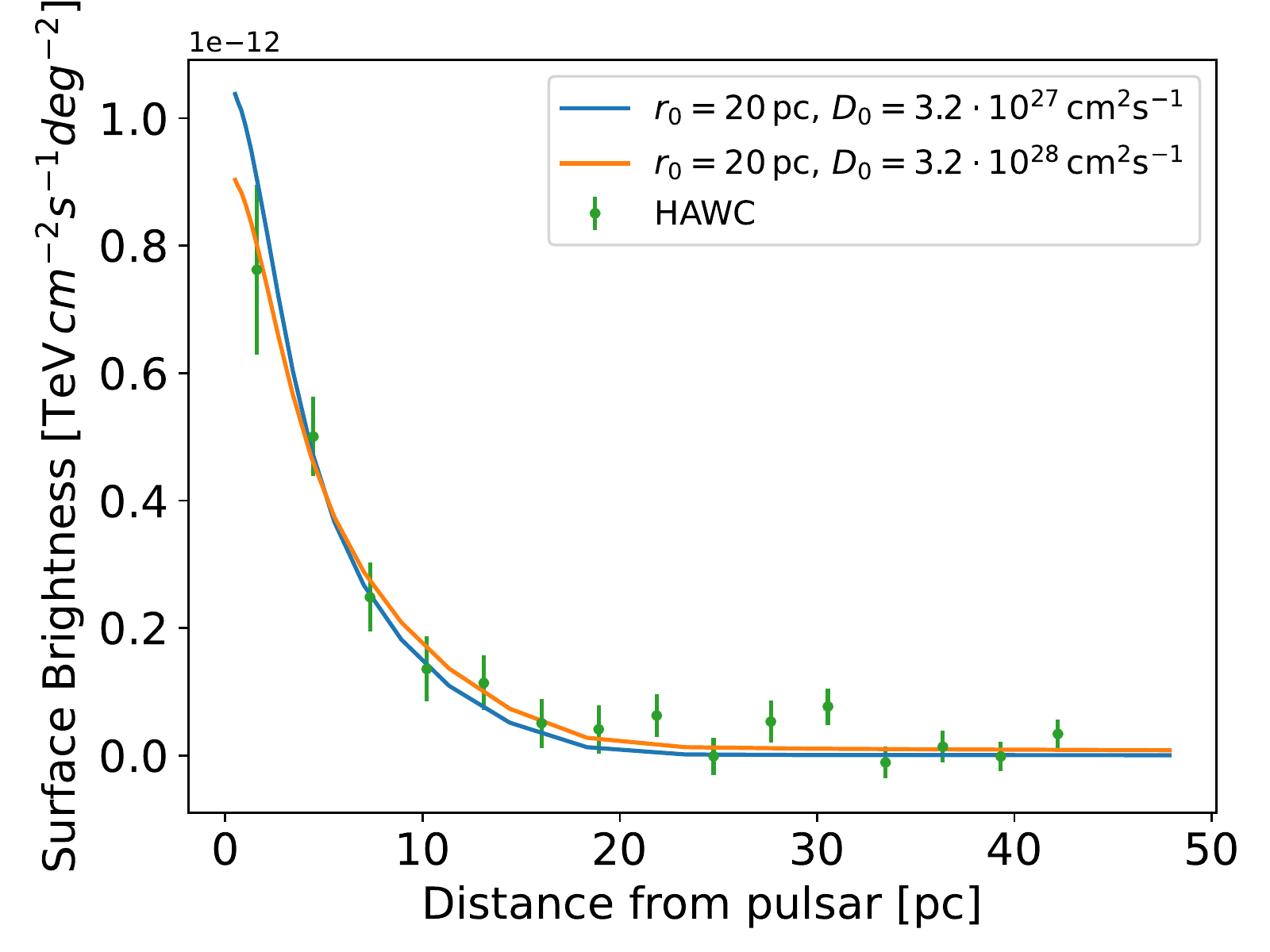}
    \caption{$\gamma$-ray spectrum for a halo size of $20\,$pc and the best-fit value of $D_0$ by HAWC and $10$ times this value.}
    \label{fig:spat_halosize}
\end{figure}

The size $r_0$ of the region where the diffusion coefficient is assumed to be suppressed may also affect the morphology of the gamma-ray emission. As discussed above, the best fit to the HAWC data requires $\lambda_{50TeV}\approx 12\,$pc, so that as long as $r_0\gg \lambda_{50TeV}$, the size of the region does not change the description of the data, and typical values of $r_0$ ranging from $30\,$pc to $120\,$pc have been used in the literature, with no appreciable change in the inferred value of $D_0$. On the other hand, 
when $r_0$ and $\lambda_{50TeV}$ become comparable, the reduced gamma-ray flux may be due to the escape of particles into the ISM at $r_0$ rather than due to energy losses. As a result the spatial profile becomes only weakly dependent on the loss length (hence on $D_0$ and $B$). 
The effect of this is illustrated by case V, shown in figure \ref{fig:spat_halosize}, where a similar spatial profile is obtained with $r_0=20\,$pc and a diffusion coefficient that is $10$ times larger than the best-fit value quoted by HAWC. For comparison, the solution with $r_0=20\,$pc and with the best-fit value by HAWC, i.e., $D_0=3.2\cdot 10^{27}\,$cm$^2$s$^{-1}$, case IV, is shown. 

Clearly these cases are very different in terms of efficiency of injection of pairs at the PWN: in fact, since the analytical solution of the problem shows that the emission is roughly degenerate with the ratio $\frac{\eta}{D}$, this translates to an upper limit on $D$ if to reproduce the same spatial profile, when the morphology is dominated by escape rather than energy losses.

For example, the case shown in fig. \ref{fig:spat_halosize} requires an efficiency of $80\%$, $E_b=1\,$TeV and $\gamma_H=2$ which is the best case scenario in terms of efficiency, i.e.,  the case with the highest normalization at $50\,$TeV energies, due to a break at high energy and a hard spectrum above the break.
These considerations show that the halo size around Geminga, based on the TeV data alone, is at least $\sim 20\,$pc large, as smaller sizes would both fail to reproduce the spatial profile of the data and require efficiencies above $100\%$.

\begin{figure}
    \centering
    \includegraphics[width=0.48\textwidth]{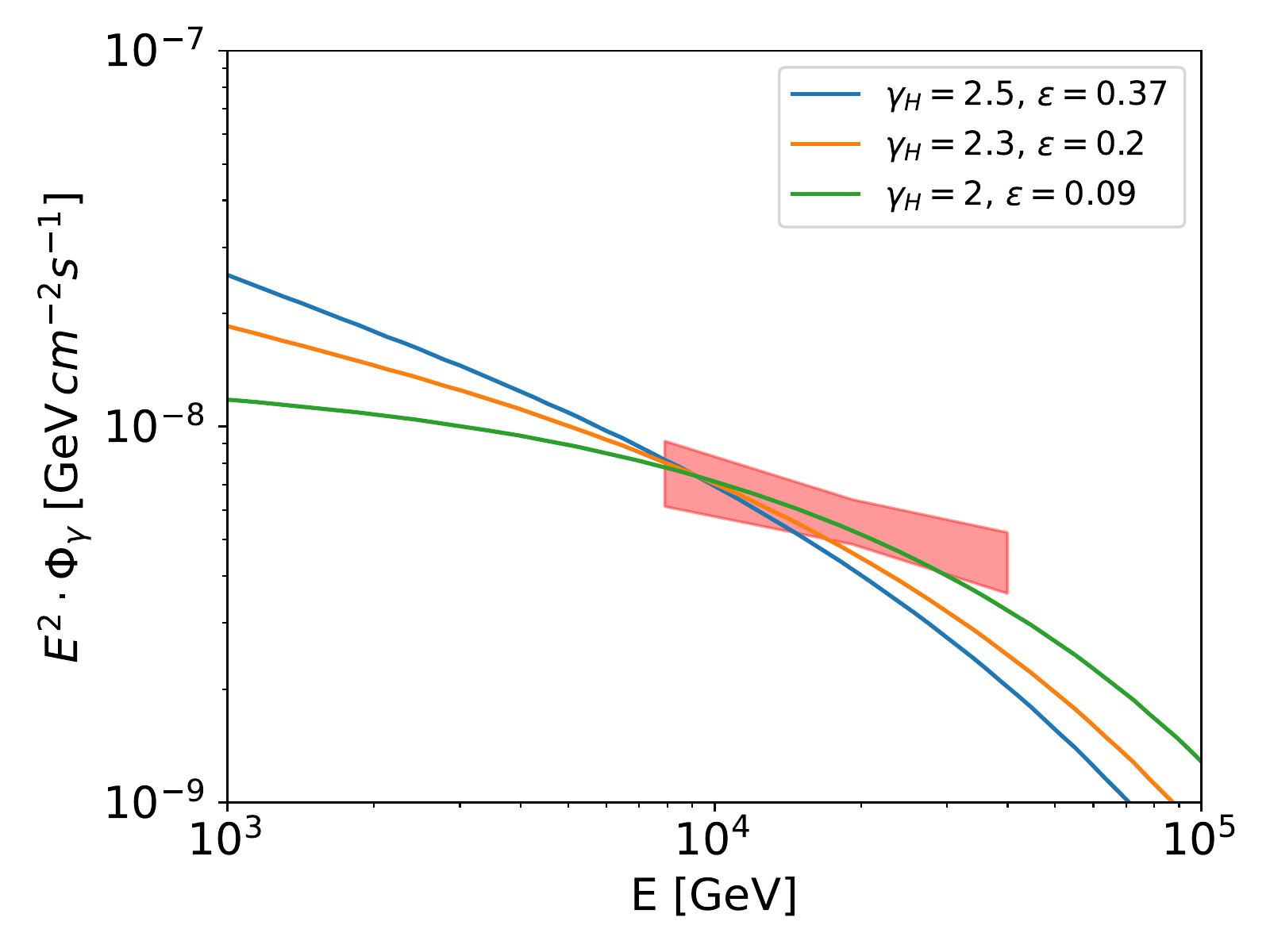}
    \includegraphics[width=0.48\textwidth]{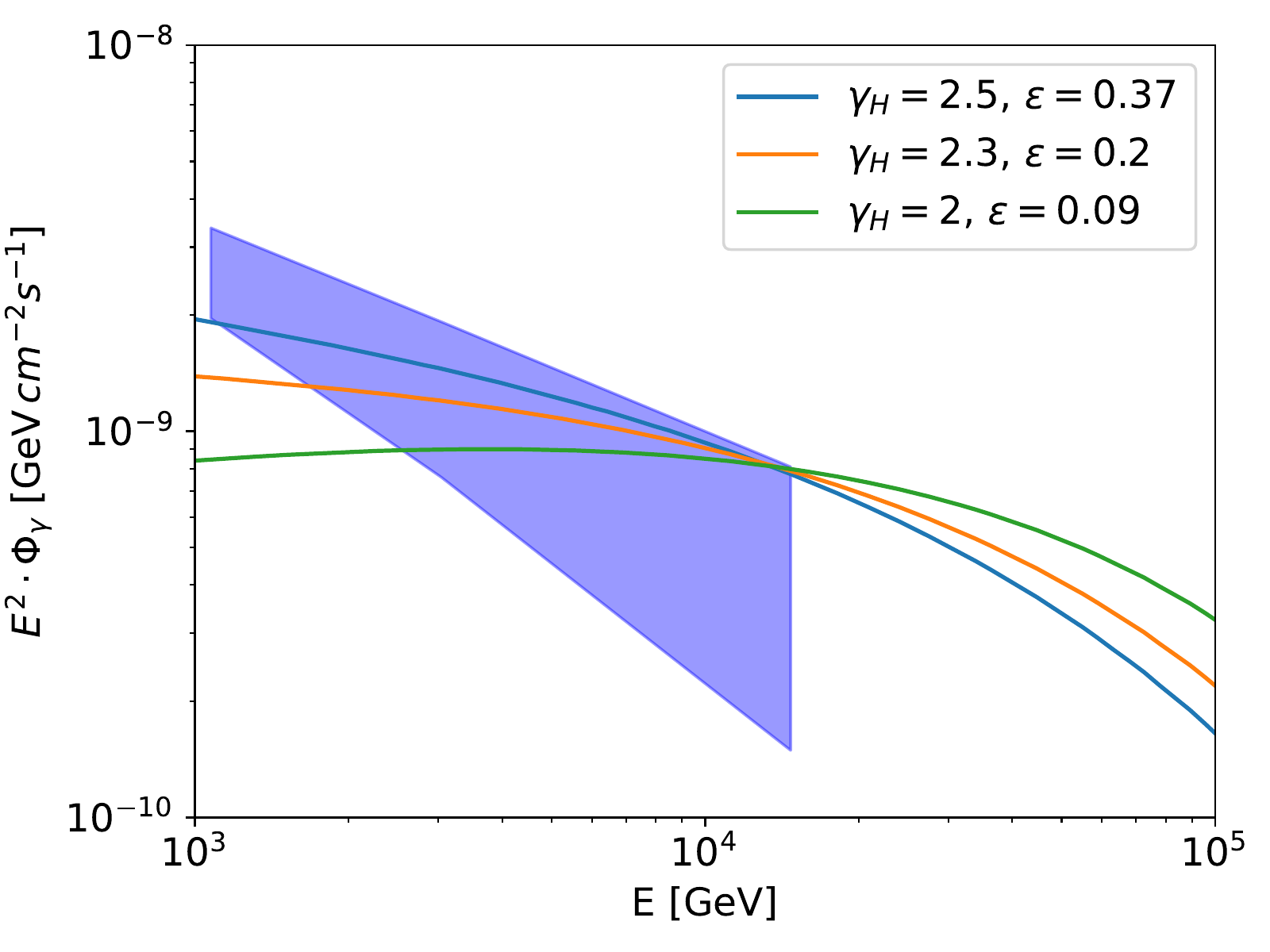}
    \caption{$\gamma$-ray spectrum produced by the released particle spectra for different injection spectra compared to the HAWC (left panel) and H.E.S.S. (right panel) measurements. All of them fit the spatial morphology of the HAWC emission around Geminga.}
    \label{fig:diff_inj}
\end{figure}

\subsection{Total Flux}

In addition to the morphology of the gamma-ray emission, its total spectrum, when available, also contains valuable information on the injection and transport of pairs. In particular, the energy dependence of the gamma-ray flux constrains the spectrum and production efficiency of pairs at the central PWN. 

In the left panel of Fig. \ref{fig:diff_inj}, we show the predicted gamma-ray spectrum for cases I, VI and VII with different values of the slope of the injected spectrum, at energies above the break. The corresponding efficiencies are also listed. The shaded area illustrates the uncertainty in the HAWC observation of the spectrum. Notice that the three curves refer to cases that fit well the morphology of the gamma-ray emission, but they all provide a rather poor description of the observed spectrum~(see also~\cite{2022PhRvD.106l3017F}). The only case that seems to be in sufficiently good agreement with the data is the one corresponding to $\gamma_H=2$, which also requires a small efficiency ($\eta=9\%$). 

In the final stages of preparation of this article, the H.E.S.S. collaboration released new measurements of the gamma-ray emission from the region of size $1^\circ$ around Geminga \cite{geminga_hess}, an appreciably smaller region than the one observed by HAWC. These data extended towards lower energies compared with HAWC data. In the right panel of Fig.~\ref{fig:diff_inj}, we compare the flux measured by H.E.S.S. with our predicted fluxes limited to the same region, for the different injection slopes quoted above. This preliminary analysis shows that the H.E.S.S. data seem to select steeper injection slopes, at odds with the HAWC data, although the large systematic uncertainties leave room for consistency between the two measurements. If taken at face value, H.E.S.S. data can be accommodated in alternative models (see e.g. \cite{geminga_hess_fang}).

\begin{figure}
    \centering
    \includegraphics[width=0.5\textwidth]{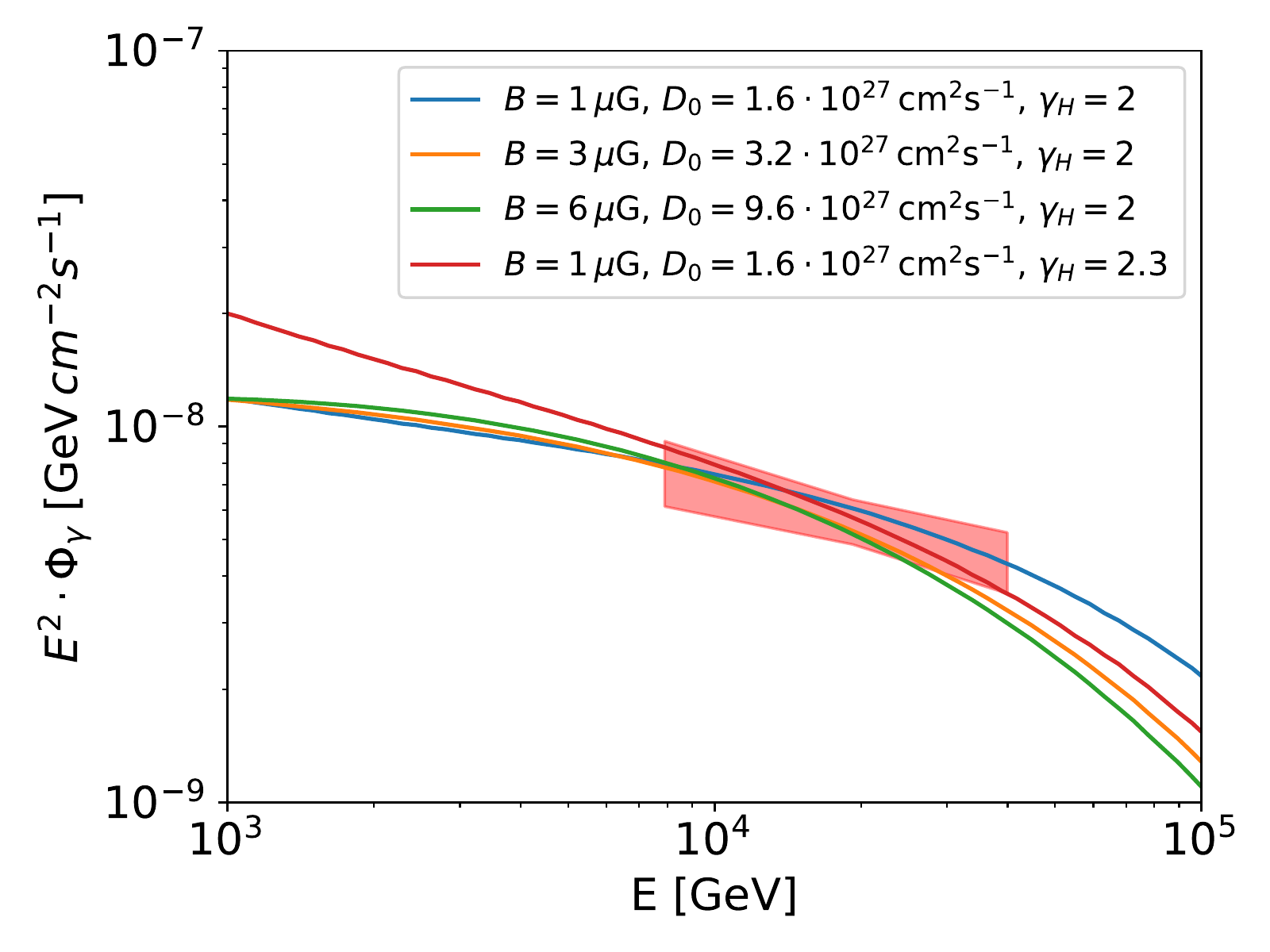}
    \caption{$\gamma$-ray spectrum produced by the released particle spectra for different magnetic field strengths with $\gamma_H=2$ and for low magnetic field and $\gamma_H=2.3$. All of them fit the spatial morphology of the HAWC emission around Geminga.}
    \label{fig:tot_flux_diff_B}
\end{figure}

In the following we will focus on the HAWC data alone and the situation illustrated in the left panel of Fig. \ref{fig:diff_inj}. The difficulties discussed above in connection with the cases listed above, are considerably reduced if a low magnetic field ($B\sim 1\mu\rm G$) in the region surrounding the PWN is adopted. This is illustrated in Fig. \ref{fig:tot_flux_diff_B} for the case of $\gamma_H=2$, where the magnetic field is allowed to vary between $1$ and $6~\mu\rm G$ (cases I-III): as discussed above, low values of the magnetic field strength imply that small diffusion coefficients must be adopted, but both the morphology and the gamma-ray flux of the halo are well reproduced if the field is low. The better agreement with the spectrum is due to the fact that small fields imply less severe energy losses and a correspondingly larger flux of high energy pairs which contribute to the gamma-ray flux observed by HAWC. In fact, the excellent agreement between the predicted and the measured spectrum led us to attempt an even larger value of $\gamma_H$. In Fig. \ref{fig:tot_flux_diff_B} we also show case VIII with $\gamma_H=2.3$ in red, which is in line with the standard high-energy spectrum of PWNe. Although a slight deficit can be seen at the highest energies, the spectrum appears to be in good agreement with the HAWC observation, and requires an efficiency $\eta\sim 9\%$. Even smaller values of the magnetic field would not automatically allow for even steeper high-energy spectra of the pairs, because at some point synchrotron losses become negligible compared to ICS, even after accounting for the Klein-Nishina suppression in the relevant energy range. 

The fact that low magnetic fields and relatively steeper spectra of the pairs seem to provide a better description of the spectrum of the gamma-ray emission as measured by HAWC also sits well with an independent measurement of the magnetic field in the region around Geminga, based on X-ray observations, which suggests $B\sim 0.8~\mu\rm G$ \cite{Geminga_xray}.

\subsection{Protons} \label{sec:protons}

As discussed above, the possibility that protons (or nuclei) could be extracted from the surface of a rapidly spinning neutron star has been discussed several times in the recent literature \cite{AmatoArons,BlasiEpsteinOlinto,Arons2003,pulsar_protons_kotera}, but no clear evidence of this phenomenon has been found insofar. The injection of protons from the PWN into the region of suppressed diffusivity would lead to an accumulation of these particles in such a region, with no appreciable energy losses, and lead to gamma-ray production through inelastic scattering. 

As discussed in Sec. \ref{sec:protons}, the spectrum of protons can be very hard, so that gamma rays of hadronic origin should be expected, if any, at the highest energies. The parameter that regulates such a contribution is the product of efficiency of proton production and gas density, $n_H$, in the region surrounding the PWN. The latter represents the target for $pp$ collisions. 



From the point of view of gamma-ray production in the $\sim 10$ TeV energy range, leptons and protons behave in quite a different way: first, leptons lose energy very quicky, on typical times of order $\sim 16\,$kyrs, so that only leptons produced recently by the PWN contribute to the gamma-ray production. On the other hand, protons are simply accumulated inside the region of reduced diffusivity until they reach the edge of the halo and escape into the ISM. In this sense, the hadronic contribution to the gamma-ray emission is more sensitive to the size $r_0$ of the halo region and to the gas density in the same region. Second, the spectrum of leptons and hadrons is quite different, and as discussed above, the hadronic contribution can only be present at high energies. 

\begin{figure}
    \centering
    \includegraphics[width=0.5\textwidth]{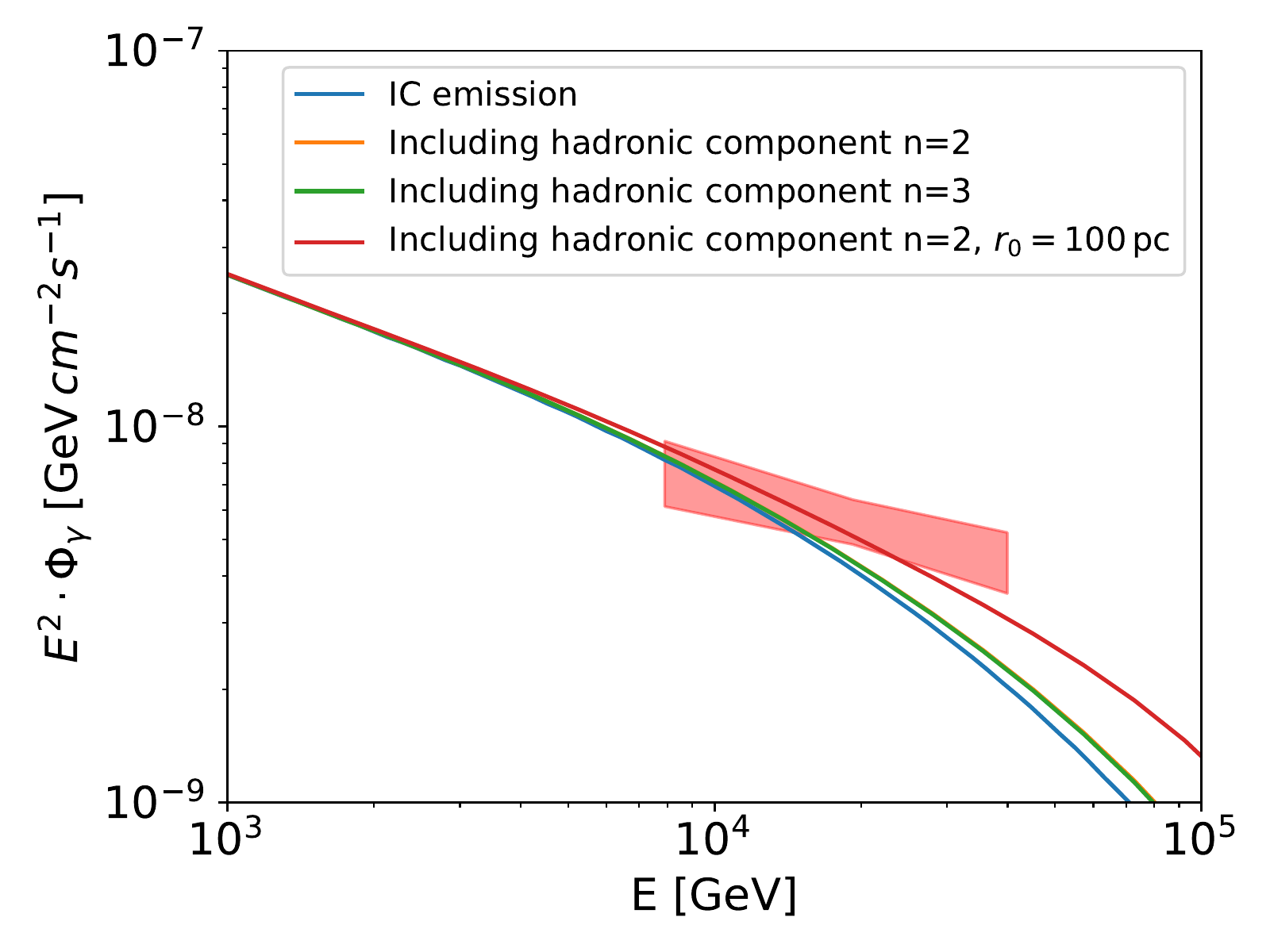}
    \caption{$\gamma$-ray spectrum produced by the released pairs including a hadronic component due to protons with different braking index $n$ and different halo size $r_0$.}
    \label{fig:protons}
\end{figure}

In fig. \ref{fig:protons}, we show the ICS contribution to the gamma-ray spectrum (solid blue curve) for parameter values of case VII in table~\ref{tab:cases}. The total gamma-ray production, including the hadronic contribution is shown in fig. \ref{fig:protons} for $r_0=50$ pc and two values of the braking index of the pulsar, $n=3$ (orange line) and $n=2$ (green line). In both cases, the hadronic contribution appears only at high energies and it is limited to a few percent of the total flux. The fluxes that include the hadronic contribution should be considered as absolute upper limits in that the efficiency of conversion of spin-down energy to protons has been maximized to unity. Moreover, the gas density has been assumed to be $n_H=1\,\rm cm^{-3}$, while the real density might be lower than that, implying a lower contribution to the gamma-ray flux.

Increasing the size $r_0$ of the region of reduced diffusivity results in a longer confinement time of protons and a correspondingly larger contribution to the total gamma-ray flux (see red curve for $r_0=100$ pc and $n=2$). 
Similarly, reducing the value of the magnetic field in the region results in a lower diffusion coefficient and therefore a larger hadronic contribution. 
This effect can amplify the hadronic component by an additional factor of $\sim 2$, making steeper spectra of $\gamma_H=2.3$ even easier to reconcile with the data.

\subsection{Escape Flux and Electron-Positron Fraction}

One of the main reasons why TeV halos have attracted so much attention is the potentially crucial effect they may have on the flux of positrons released into the ISM as CRs. Initial reactions to the discovery of the extended emission around Geminga were based on the assumption that the diffusion coefficient in the Galaxy may be smaller than deduced based on the measured Secondary/Primary ratios of the flux of nuclei, everywhere in the Galaxy \cite{geminga_hawc}. In this situation, it would be difficult to expect to receive positrons from PWNe in the form of CRs, and the observed positron excess would require an alternative explanation. The assumption that the diffusion coefficient is suppressed everywhere is clearly extreme and not necessary. In fact, it is more natural to expect that regions of suppressed diffusion may exist near sources \cite{2019Galaxy} for a variety of reasons. One could indeed reverse the argument and interpret the detection of TeV halos around PWNe as the clearest evidence that PWNe are indeed sources of CR electron-positron pairs \cite{2017Hooper}. 

Even though the suppression of CR diffusivity is a local phenomenon, it may still severely affect the spectrum of pairs released into the ISM, thereby changing the connection between the positron (and electron) spectra observed at the Earth and those inferred in PWNe. In fact, this modification might even be required by existing data, as recently discussed in \cite{e_factories}: the authors find that, in order to reproduce the positron CR spectrum and the positron ratio at high energies, the best-fit injection spectrum of pairs from PWNe into the ISM (slope $\sim 2.8$) is appreciably steeper than typically found in X-ray and gamma-ray observations of these sources ($\gamma_H\sim 2.3-2.5$). 

In this section, we explore the possibility that this requirement may reflect the effect of energy losses inside the regions of suppressed diffusivity. In order to do so, we determine the escape flux from such regions and interpret it as the effective spectrum injected by the source into the ISM. The calculation is specialized to the region of reduced diffusivity around the Geminga PWN.

\begin{figure}
    \centering
    \includegraphics[width=0.5\textwidth]{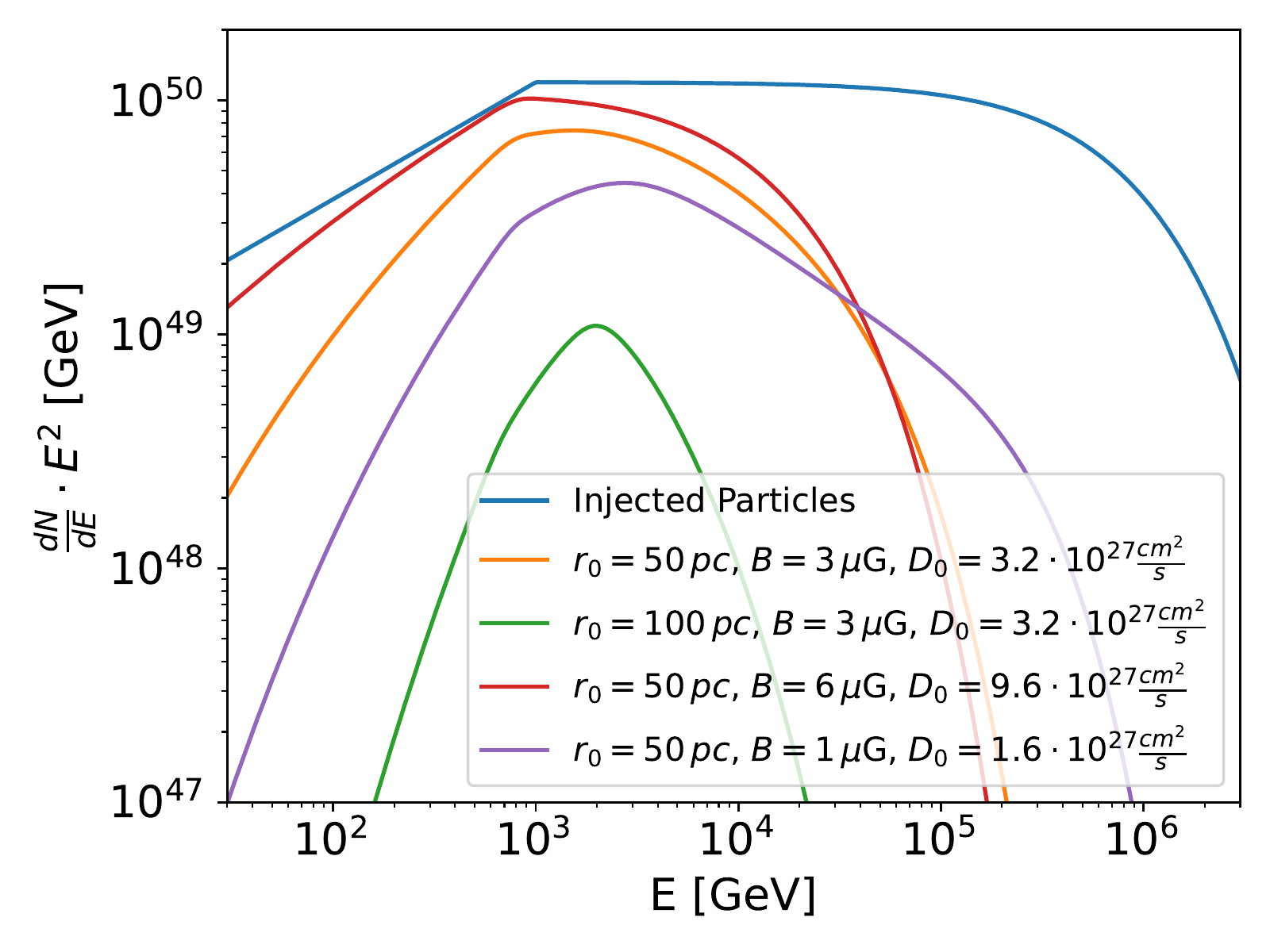}
    \caption{Time-integrated escape flux from the halo and injection for different cases of magnetic field $B$, diffusion coefficient $D_0$ and halo size $r_0$. The injection efficiency for all cases is set to $1$ to give a meaningful comparison between the different cases with the respective injection.}
    \label{fig:esc_flux}
\end{figure}

In Fig. \ref{fig:esc_flux} we show the spectrum of pairs contributed by the PWN (broken solid blue line), as described by eq. \ref{eq:injection}, with $\gamma_H=2$, and the time-integrated escape flux for different values of $r_0$ and of the magnetic field in the region around the PWN. The integration in time is carried out from the time when the pulsar leaves the remnant to the current age of the pulsar. One can see that, depending on the value of the diffusion coefficient and of the halo size, even after $342\,$kyrs only a fraction of particles manage to escape the halo. In the considered energy range from $10\,$GeV to $10\,$PeV this fraction is about $20\%$ for $r_0=50\,$pc and $D_0=3.2\cdot 10^{27}$cm$^2$s$^{-1}$, meaning that most particles, especially at lower energies, are still trapped within $50\,$pc of the source.

The probably most important effect emphasized by Fig. \ref{fig:esc_flux} is that the spectrum of pairs is severely affected by energy losses during the confinement time: the time-integrated escape flux is made steeper at the highest energies by ICS and synchrotron losses, with a pronounced cutoff at energies much lower than the potential drop of the pulsar.

In the case of low magnetic field strength, $B=1\,\mu$G, the spectrum of escaped pairs is steeper than what is injected at the termination shock over at least an order of magnitude in energy. 
This finding might explain the observation described above that in order to explain the positron fraction, rather steep spectra released into the ISM with an average index of $2.8$ are needed \cite{e_factories}. Clearly, for these modifications to be attributed to the existence of regions of small diffusivity around PWNe, it is necessary that TeV halos are a rather generic phenomenon, present around most PWNe. 

At lower energies the spectrum is modified mainly because pairs do not have enough time, within the age of the pulsar, to escape the region of reduced diffusivity. 

\begin{figure}
    \centering
    \includegraphics[width=0.5\textwidth]{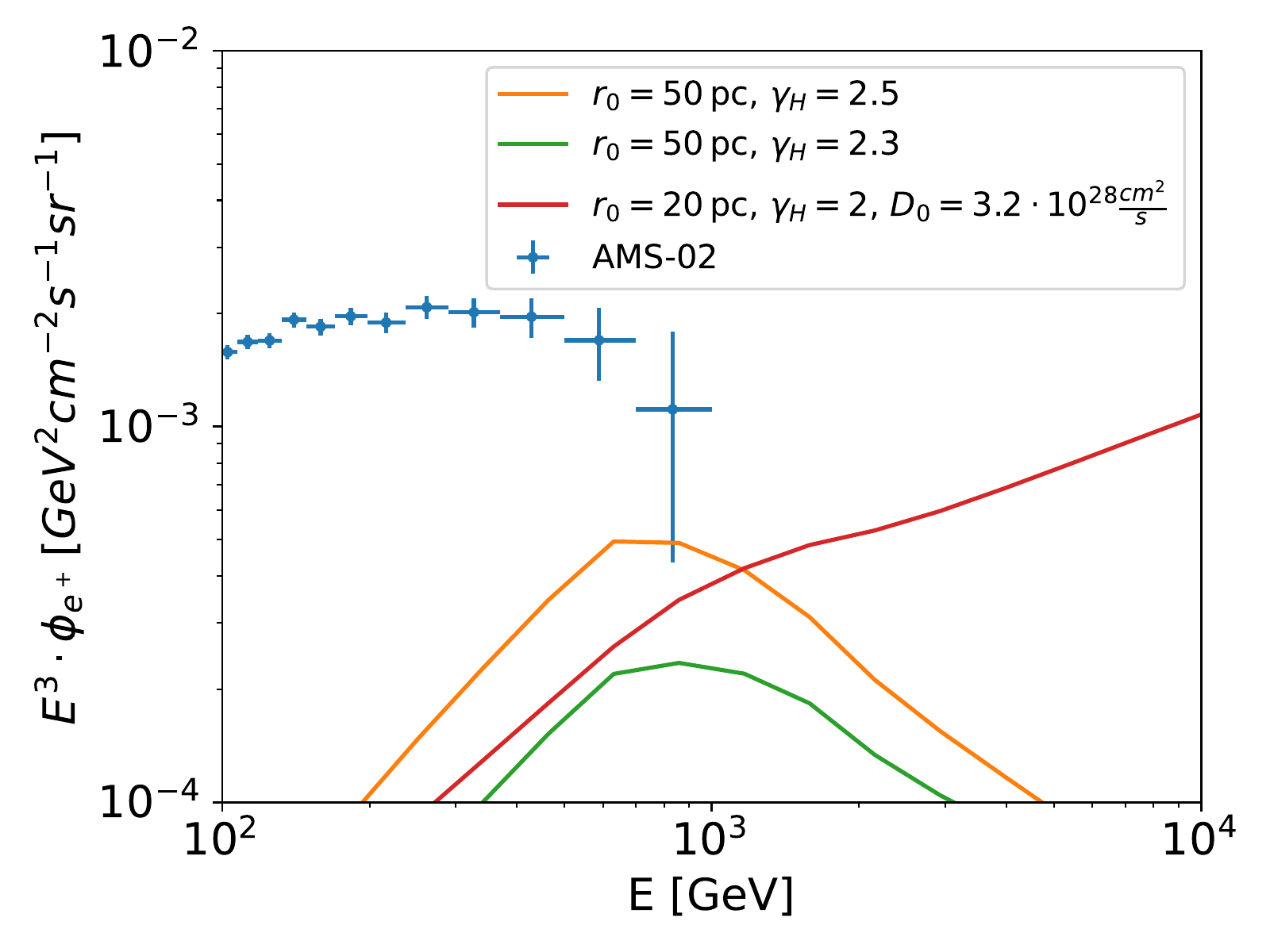}
    \caption{Expected contribution from the Geminga PWN to the local positron flux for different cases that fit the spatial morphology and total flux measured by HAWC.}
    \label{fig:e+_flux}
\end{figure}

Given the close distance to Geminga, it is important to check what would be its contribution for the positron flux at the location of the Earth, taking into account the effect of energy losses in the TeV halo region. We choose three cases which fit well the morphology of the TeV emission observed by HAWC and we show the positron flux at Earth in fig. \ref{fig:e+_flux}. The adopted halo parameters are listed in the figure.

The fluxes are calculated with the two-zone model assuming a Kolmogorov scaling of the diffusion coefficient and normalizing it in the ISM to the one found in \cite{BC_diff} at $100\,$TeV. The injection efficiencies for the different cases are fixed according to the analysis above, to fit the spatial profile and total flux measured by HAWC. It is clear that the fluxes vary significantly between the different cases, due to the role of energy losses. Although all cases explain the spatial profile of the TeV $\gamma$-ray emission around Geminga, the positron flux at Earth varies by more than one order of magnitude.

The values obtained for a fixed halo size of $50\,$pc range from $50\%$ ($\gamma_H=2.5$) down to $5\%$ ($\gamma_H=2$, not shown) of the positron flux at 1 TeV depending on the injection slope.
Furthermore, a smaller halo size would favor the escape of high-energy particles, thereby increasing the positron flux at high energies.

\section{Conclusions}\label{sec:conclusion}

The recent discovery of the so-called TeV halos, regions of diffuse TeV gamma-ray emission around selected PWNe, has stimulated a burst of interest in its potential implications mainly in terms of the origin of the positron excess. 

An issue that is probably been less appreciated in the community is that of establishing the physical reason for the existence of regions of suppressed diffusivity and their possible connection with sources of high-energy particles. The first step in the investigation of this issue is the characterization of these halos, namely the dependence of the inferred properties on the parameters of the problem, for instance, the size of these regions of reduced diffusivity, the strength of the magnetic field in such regions and the spectrum of the pairs released by a PWN into the ISM. All these parameters define both the morphology and the spectrum of the gamma-ray emission, which in turn can be used to constrain the properties of TeV halos and their origin. 

Our investigation confirms that the diffusivity of CRs in the region around Geminga has to be suppressed by a factor $\sim 1000$ \citep{geminga_hawc,halos_DM,tang_piran}. However, we also find that a relatively small halo, of size $\sim 20$ pc, may also be compatible with the observations, while being somewhat less demanding in terms of suppression of the diffusion coefficient. 

The spectrum of the observed TeV halo around Geminga seems to require a rather hard injection spectrum of pairs in the $\gtrsim 50$ TeV energy region, with slope $\gamma_H\sim 2$, at odds with the spectra that are often adopted and inferred from multi-frequency observations of PWNe. The exception to this finding is the case in which the magnetic field in the halo region is low, $B\sim 1\mu G$, which seems to also be consistent with X-ray observations \cite{Geminga_xray}. Such a low field reduces the effect of energy losses of high energy pairs, thereby allowing for steeper spectra, while requiring relatively low efficiency, $\sim 9\%$, in terms of conversion of spin-down luminosity into pairs. However, this case requires a severe suppression of the diffusion coefficient around Geminga, compatible with the one initially inferred by the HAWC collaboration \cite{geminga_hawc}. 
Furthermore, new measurements by H.E.S.S. \cite{geminga_hess} seem to prefer steeper spectra, which would make the injected spectrum more similar to the one of other PWNe. However, the large experimental uncertainties make it difficult to reach a firm conclusion about the slope of the spectrum of pairs injected into the ISM.

We also calculated the gamma-ray flux contributed by a yet speculative proton population of particles extracted from the neutron star surface. This contribution, which could potentially become relevant at the highest gamma-ray energies, due to inelastic collisions of protons with energy close to the potential drop of the pulsar, is however found to be subdominant, representing at most a few percent of the total measured gamma-ray flux. This general conclusion applies to halos with size $\lesssim 50$ pc and with braking index of the pulsar in the standard range $n=2-3$. On the other hand, smaller diffusion coefficients (tied to lower magnetic fields) and/or larger sizes of the region of reduced diffusivity, $r_0\sim 100$ pc, would enhance the role of protons in that they are confined for longer times (proportional to $r_0^2/D$) while not losing appreciable energy. Such halo sizes would, however, have rather serious implications in terms of CR and positron transport at large, since it would imply that almost everywhere in the disc of the Galaxy the diffusion coefficient is severely suppressed.  

One of the reasons why we engaged in this investigation was the recent finding \cite{e_factories} that the spectrum of pairs that PWNe are required to inject into the ISM to fit the positron spectrum and the positron fraction observed at the Earth is, at high energies ($\gtrsim 100$ GeV), much steeper ($\sim 2.8$) than the spectrum inferred from multi-frequency observations of PWNe ($\sim 2.3-2.5$). In this sense, the detection of the TeV halo around Geminga was very interesting in that it guaranteed longer confinement times of pairs around the PWN and correspondingly more severe energy losses that those suffered during propagation in the Galaxy. In other words, the effective spectrum injected by a PWN into the ISM would be the spectrum of pairs escaping the region of reduced diffusivity. 

The calculation of this effect requires the adoption of a two-zone model, in which the diffusion coefficient is different within a region of size $r_0$ around the PWN and in the Galaxy at large. We found the existing two-zone model, used for instance in \cite{tang_piran}, to be incorrect (the solution does not conserve the particle flux at the boundary between the two zones). The same mistake together with the corrected version was found in \cite{Osipov} and allows us to infer the appropriate spectrum of pairs inside the region of reduced diffusivity and the escape spectrum, to be used as an effective injection spectrum into the ISM. We found that energy losses inside the region of reduced diffusivity severely affect the spectrum of pairs. For the case with low magnetic field, $B\sim 1 \mu G$, the spectrum of pairs escaping into the ISM is steeper than the one in the PWN over more than one order of magnitude in energy. In addition, the spectrum of pairs injected into the ISM has a cutoff at energies much lower than the potential drop of the pulsar, as a result of radiative losses in the region of low diffusion coefficient. 

These results show that if these regions exist around most PWNe, the connection between the positron spectrum observed at the Earth and the one inferred from radiation in individual PWNe is not trivial and needs to take into account transport in the region of reduced diffusivity. 

Clearly, the assessment of this issue and of how common these halos are is also crucial for CR transport at large, because different models for the origin of the halos reflect in different implications in terms of grammage accumulated by CRs near sources.

By fixing the overall normalization of the injection spectra to scenarios that explain the spatial morphology and total flux measured by HAWC, we calculate the expected contribution to the local positron flux by Geminga using the correct two-zone model \cite{Osipov}. We find that the expected flux varies by more than one order of magnitude depending both on the spectral properties of the injection and the halo size.


The future detection of more TeV halos and the characterization of such diffuse emission, following the procedure illustrated here, should provide us with better clues to the origin of the regions of reduced diffusivity and will eventually result in a better understanding of the positron flux at the Earth and of CR transport in the Galaxy. 

\bibliographystyle{apsrev4-2}
%

\end{document}